\documentclass[aps,prd,draft,showpacs]{revtex4}

\begin{document}

\title{Derivation of the Recursion Relation for the Feynman Diagrams
of the CJT effective action}
\author{Chungku Kim}
\date{\today}

\begin{abstract}
We derive a new recursion relation to obtain the Feynman diagrams of the
Cornwall-Jackiw-Toumboulis(CJT) effective action by using the functional
derivative identities. By using this recursion relation we show the
two-particle-irreducibility of the Feynman diagrams of the CJT effective
action by induction. We apply the recursion
relation to obtain the Feynman diagrams of the CJT effective action up to the 
four-loop order in case of the bosonic field theory.
\end{abstract}
\maketitle


\affiliation{Department of Physics, College of Natural Science, Keimyung
University, Daegu 705-701, KOREA}

\section{Introduction}

The effective action plays an important role in studies of the vacuum
instability, the dynamical symmetry breaking and the dynamics of composite
particles\cite{Sher} for the given particle physics model. Among the several
schemes for the nonperturbative evaluation of the effective action, the CJT
effective action\cite{CJT} was widely used as a resummation scheme which
gives a self-consistent loop expansion of the effective action while
preserving the symmetry of the original theory. In case of the connected
and the one-particle-irreducible (1PI) effective action,
the recursive generation of the Feynman diagrams was obtained for the
multicomponent $\phi ^{4}$-theory, QED and the scalar QED theories \cite
{R2}\cite{R3} \cite{R4}\cite{R5}\cite{R6}\cite{R7}\cite{R8} by using the
functional integral identities $\int D\Phi \ \frac{\delta }{\delta \Phi }
F[\Phi ]=0$. In case of the CJT effective action which is the second Legendre
transformation of the generating functional, the recursion relation was
obtained by extracting the two-particle-irreducible (2PI) Feynman diagrams from 
the 1PI effective action \cite{2PI}. Since the Feynman diagrams of the
effective action obtained from the n-th Legendre transformation of the
generating functional are not always the n-particle-irreducible (nPI) Feynman
diagram \cite{kim1}, obtaining the recursion relation  
by extracting nPI Feynman diagrams is not possible when $n > 2$.

Recently, we have derived a new method to obtain the recursion relation for
the effective action by using the functional derivative identity \cite{kim2}.
This method can be extended to the case of the effective action obtained from
the n-th Legendre transformation of the generating functional and in
this paper, we apply this method to the CJT effective action. In Sec.II, we
derive the recursion relation for the CJT effective action. We prove the 
two-particle-irreducibility of the CJT effective action by using the recursion
relation. Then we apply to the case of Feynman diagrams up to the
four-loop order effective action for the bosonic field theory. In Sec.III, we
give some discussions and conclusions.

\section{A\ Recursion Relation for the Feynman Diagrams of the CJT Effective
Action}

In this section, we will first derive a recursion relation for the Feynman
diagrams of the CJT effective action for the bosonic field theory with the
classical action

\begin{equation}
S[\Phi ]=\int \{\frac{1}{2}\Phi _{A}\Delta _{AB}^{-1}\Phi _{B}+S^{int}[\Phi
]\}.
\end{equation}
where $S^{int}[\Phi ]$ contains the higher vertices which
appear in lattice regularization\cite{lattice} as well as the cubic and the
quartic interactions. In this paper, we use a notation where the capital
letters contain both the space-time variables and the internal indices and
the repeated capital letters mean both the integration over continuous
variables and the sum over internal indices. For example, if the capital
letter $A$ contains a space-time variable $x$ and the internal index $i,$

\begin{equation}
J_{A}\Phi _{A}\equiv \sum_{i}\int d^{4}xJ_{i}(x)\Phi _{i}(x).
\end{equation}
The generating functional $W[J,K]$ is given by

\begin{equation}
W[J,K]=-\hbar \ln \int D\Phi \ Exp[-\frac{1}{\hbar }(S[\Phi ]+J_{A}\Phi _{A}+%
\frac{1}{2!}K_{AB}\Phi _{A}\Phi _{B})],
\end{equation}
where the external source $K_{AB}\ $is symmetric under the exchange of the
indices. The functional derivative of the $W[J,K]$ with respect to
the external sources are given by the classical field $\phi$ and the
full propagator G as

\begin{eqnarray}
\frac{\delta W[J,K]}{\delta J_{A}} &=&\langle \Phi _{A}\rangle =\phi _{A}, \\
2\frac{\delta W[J,K]}{\delta K_{AB}} &=&\langle \Phi _{A}\Phi _{B}\rangle
=\phi _{A}\phi _{B}+\hbar G_{AB},
\end{eqnarray}
From (3) and (5) we can see that
\begin{equation}
G_{AB}\equiv -\text{ }\frac{\delta ^{2}W[J,K]}{\delta J_{A}\delta J_{B}}.
\end{equation}
By inverting (4) and (5), one can obtain the functionals $J[\phi ,G]$
and $K[\phi ,G]$. Then the CJT effective action is given by the second Legendre
transformation of the generating functional as

\begin{equation}
\Gamma [\phi ,G]=W[J,K]-J_{A}\phi _{A}-\frac{1}{2}K_{AB}(\phi _{A}\phi
_{B}+\hbar G_{AB}).
\end{equation}
From (4),(5)and (7), one can obtain the following relations :

\begin{eqnarray}
\frac{\delta \Gamma [\phi ,G]}{\delta \phi _{A}} &=&-J_{A}-K_{AB}\phi _{B},
\\
\frac{\delta \Gamma [\phi ,G]}{\delta G_{AB}} &=&-\text{ }\frac{\hbar }{2}%
K_{AB}.
\end{eqnarray}
and from (3) and (7), we obtain

\begin{equation}
\exp \{-\frac{1}{\hbar }\Gamma [\phi ,G]\}=\int D\Phi \ \text{exp}\{-\frac{1%
}{\hbar }(S(\Phi )+J_{A}(\Phi _{A}-\phi _{A})+\frac{1}{2}K_{A}(\Phi _{A}\Phi
_{B}-\phi _{A}\phi _{B}-\hbar G_{AB}))\}.
\end{equation}
By expanding the effective action $\Gamma [\phi ,G]$ around $\hbar $, we obtain

\begin{equation}
\Gamma [\phi ,G]=\sum_{l=0}\hbar ^{l}\Gamma ^{(l)}[\phi ,G],
\end{equation}
we can obtain the loop-wise expansion of $\Gamma [\phi ,G]$\cite{Coleman}.
The first two terms are given by\cite{CJT} 
\begin{equation}
\Gamma ^{(0)}[\phi ,G]=S[\phi ]\text{, }\Gamma ^{(1)}[\phi ,G]=\frac{1}{2}%
Tr\ln G^{-1}-\frac{1}{2}TrG(G^{-1}-D^{-1}).
\end{equation}
where 
\begin{equation}
D_{AB}^{-1}[\phi ]\equiv \frac{\delta ^{2}S[\phi ]}{\delta \phi _{A}\delta
\phi _{B}}.
\end{equation}

Now consider the functional identities satisfied by the two sources $J[\phi
,G]$ and $K[\phi ,G]$

\begin{equation}
\frac{\delta J_{A}}{\delta \phi _{C}}\frac{\delta \phi _{C}}{\delta J_{B}}+%
\frac{\delta J_{A}}{\delta G_{CD}}\frac{\delta G_{CD}}{\delta J_{B}}=\delta
_{AB}
\end{equation}
and 
\begin{equation}
\frac{\delta K_{AE}}{\delta \phi _{C}}\frac{\delta \phi _{C}}{\delta J_{B}}+%
\frac{\delta K_{AE}}{\delta G_{CD}}\frac{\delta G_{CD}}{\delta J_{B}}=0.
\end{equation}
By eliminating the term $\frac{\delta G_{CD}}{\delta J_{B}}$ from (14) and
(15), we obtain 
\begin{equation}
(\frac{\delta J_{A}}{\delta \phi _{C}}-\frac{\delta J_{A}}{\delta G_{DE}}%
\Delta _{DE,PQ}^{-1}\frac{\delta K_{PQ}}{\delta \phi _{C}})\frac{\delta \phi
_{C}}{\delta J_{B}}=\delta _{AB}
\end{equation}
where 
\begin{equation}
\Delta _{AB,CD}\equiv \frac{\delta K_{AB}}{\delta G_{CD}}=-\frac{2}{\hbar }%
\frac{\delta ^{2}\Gamma }{\delta G_{AB}\delta G_{CD}}\text{ }
\end{equation}
We have used (9) in deriving the last line of the above equation. From (4) and (6),
we can obtain

\begin{equation}
\frac{\delta \phi _{C}}{\delta J_{B}}=\frac{\delta ^{2}W[J]}{\delta
J_{C}\delta J_{B}}=-G_{CB}
\end{equation}
From (8) and (9), we can obtain 
\begin{equation}
J_{A}=-\frac{\delta \Gamma [\phi ,G]}{\delta \phi _{A}}+\frac{2}{\hbar }%
\frac{\delta \Gamma [\phi ,G]}{\delta G_{AB}}\phi _{B},
\end{equation}
By substituting (9),(18) and (19) into (16), we can obtain 
\begin{equation}
-\text{ }\frac{\delta ^{2}\Gamma }{\delta \phi _{A}\delta \phi _{B}}+\frac{2%
}{\hbar }\frac{\delta \Gamma }{\delta G_{AC}}-2\hbar \gamma _{A,DE}\Delta
_{DE,PQ}^{-1}\gamma _{B,PQ}=-G_{AB}^{-1}
\end{equation}
where 
\begin{equation}
\gamma _{A,BC}\equiv \frac{1}{\hbar }\frac{\delta ^{2}\Gamma }{\delta \phi
_{A}\delta G_{DE}}
\end{equation}
Since $\Gamma ^{(0)}$ does not depend on $G$, $\Delta $ and $\gamma $ can be
expanded around $\hbar $ as

\begin{equation}
\Delta _{AB,CD}=\sum_{l=0}\hbar ^{l}\Delta _{AB,CD}^{(l)}\text{ and }\gamma
_{A,BC}=\sum_{l=0}\hbar ^{l}\gamma _{A,BC}^{(l)}
\end{equation}
where 
\begin{equation}
\Delta _{AB,CD}^{(l)}=-2\frac{\delta ^{2}\Gamma ^{(l+1)}}{\delta
G_{AB}\delta G_{CD}}\text{ and }\gamma _{A,BC}^{(l)}=\frac{\delta ^{2}\Gamma
^{(l+1)}}{\delta \phi _{A}\delta G_{DE}}.
\end{equation}
From (12), we can obtain 
\begin{equation}
\Delta _{AB,CD}^{(0)}=-G_{AC}^{-1}G_{BD}^{-1},\text{ }\Delta
_{AB,CD}^{-1(0)}=-\text{ }G_{AC}G_{BD}\text{ and }\gamma _{A,BC}^{(0)}=\frac{%
1}{2}S_{ABC}[\phi ]
\end{equation}
where 
\begin{equation}
S_{A_{1\cdot \cdot \cdot }A_{N}}[\phi ]\equiv \frac{\delta ^{N}S[\phi ]}{%
\delta \phi _{A_{1}}\cdot \cdot \cdot \delta \phi _{A_{N}}}.
\end{equation}
Now, by using (12), we can check that order $\hbar ^{0}$ term of (20) is
satisfied and the higher order terms are given by
\begin{equation}
G_{AB}\frac{\delta \Gamma ^{(l)}[\phi ,G]}{\delta G_{AB}}=\text{ }\frac{1}{2}%
G_{AB}\frac{\delta ^{2}\Gamma ^{(l-1)}[\phi ,G]}{\delta \phi _{A}\delta \phi
_{B}}+\sum_{{{p,q,r}}\atop{{p+q+r=l-2}}}G_{AB}\gamma _{A,DE}^{(p)}\Delta
_{DE,PQ}^{-1(q)}\gamma _{B,PQ}^{(r)}\text{ ( }l\geq 2\text{ )}
\end{equation}
Note that the result of the operation $G_{AB}\frac{\delta \Gamma ^{(l)}}{%
\delta G_{AB}}$ is equal to multiplying each diagrams in $\Gamma ^{(l)}$ by
the number of its propagators $G$. Eq.(26) is the central result of this
paper. 

Before applying (26) to specific examples, let us show
the two-particle-irreducibility of the Feynman diagrams of $\Gamma ^{(l)}[\phi ,G]$
by induction. For this purpose, let us assume
that all the Feynman diagrams of the $\Gamma ^{(k)}[\phi,G] (k<l)$
are two-particle-irreducible. Since the operation $\frac{\delta }{\delta \phi }$
which appears at the first term of right hand side (RHS) of (26)
changes only the vertex factors of the $\Gamma ^{(l-1)} [\phi ,G]$ which have 2PI
Feynman diagrams by assumption ( see (32) and the second column
of the Table 1 ), the first term of right hand 
side (RHS) of (26) consists of 2PI Feynman diagrams. 
The graphical representation of the second term of the RHS of (26) is given in Fig.1 and
let us note the following properties of the $\gamma^{(k)} $ and $\Delta^{-1(k)} 
(k \leq l-2)$ which contribute to $\Gamma ^{(l)} [\phi ,G]$.
\newline
\newline
(A) $\gamma_{A,BC}^{(k)} (k \leq l-2)$ has a one-particle-irreducibility so that it can
not be divided by two separate
parts by cutting any one of the lines (see (34) and the third column of Table 1).
This follows from (23), where the operation $\frac{\delta }{\delta \phi }$ changes the
vertex factors and the operation $\frac{\delta }{\delta G}$ removes only one
line from $\Gamma^{(k+1)}$ which have the two-particle-irreducible Feynman diagrams by
assumption. 
\newline
\newline
(B) $\Delta_{AB,CD}^{-1(k)} (k \leq l-2)$ does not have a diagram of the type given
in Fig.2 so that it can not be divided by two separate parts containing the index AB
and CD respectively. In order to see this, first note that $\Delta _{AB,CD}^{(k)}
 (k \leq l-2)$ does not have a diagram 
of the type given in Fig.2. This follows from (23) such that if we 
connect the points $A$ and $B$  of $\Delta_{AB,CD}^{(k)}$ with the propagator $G_{AB}$ and 
connect the points $C$ and $D$ with the propagator $G_{CD}$ , then we should
obtain the 2PI diagram. If $\Delta_{AB,CD}^{(k)}$ have a diagram of the type given in Fig.2,
this is impossible and hence $\Delta _{AB,CD}^{(k)}$ does not have a diagram shown
in Fig.2 (see (35) and the fourth column of Table 1 ). Now, in order to see that
$\Delta _{AB,CD}^{-1(k)}$ does not have a
diagram of the type given in Fig.2 by induction, let us assume that $\Delta
_{RS,CD}^{-1(m)}(m \leq k-1)$ does not have a diagram of the type given in Fig.2.
By using (24), the perturbative expansion of the $\Delta_{AB,CD}^{-1(k)}$ is given by 
\begin{equation}
\Delta _{AB,CD}^{-1(k)}=-\text{ }G_{AP}G_{BQ}(\Delta _{PQ,RS}^{(1)}\Delta
_{RS,CD}^{-1(k-1)}+\Delta _{PQ,RS}^{(2)}\Delta _{RS,CD}^{-1(k-2)}+...+\Delta
_{PQ,RS}^{(k)}\Delta _{RS,CD}^{-1(0)}).
\end{equation}  
Since both the $\Delta _{AB,CD}^{(m)}(m \leq k)$ and $\Delta
_{AB,CD}^{-1(k)} (m \leq k-1)$ of RHS of the above equation do not have a diagram
of the type given in Fig.2, $\Delta_{AB,CD}^{-1(l)}$
does not have a diagram of the type given in Fig.2 also (see (36)).
\newline
\newline
From the properties (A) and (B) given above, it is clear that one can not
divide the second term of RHS of (26) given in Fig.1 by two separated parts by
cutting any two lines of the propagators. This completes the proof that all 
the Feynman diagrams of $\Gamma ^{(l)}[\phi ,G]$ are two-particle-irreducible.

Now, consider the $l=2$ case of (26). In this case, the first term of (26)
becomes 
\begin{equation}
\frac{1}{2}G_{AB}\frac{\delta ^{2}\Gamma ^{(1)}[\phi ,G]}{\delta \phi
_{A}\delta \phi _{B}}=\frac{1}{4}S_{ABCD}[\phi ]G_{AB}G_{CD}
\end{equation}
and the second term of (26) becomes

\begin{equation}
G_{AB}\gamma _{A,DE}^{(0)}\Delta _{DE,PQ}^{-1(0)}\gamma _{B,PQ}^{(0)}=-\frac{%
1}{4}S_{ADE}[\phi ]S_{BPQ}[\phi ]G_{AB}G_{DP}G_{EQ}
\end{equation}
In this paper we use the graphical representation where a line represents the propagator
$G$ and a $ n-$point vertex have the factor $S_{A_{1\cdot \cdot \cdot }A_{n}}$. Also a
box with an capital letter represents the vertex which have indices that is
not contracted with the propagators attached to it so that 
\begin{equation}
\begin{picture}(130,40) \put(10,21){\framebox(10,10){}} \put(6,35){$A..B$}
\put(12,21){\line(-1,-1){10}}\put(15,21){\line(0,-1){13}}\put(20,21){%
\line(1,-1){10}} \put(18,15){$..$} \put(45,23){$=$}
\put(0,2){$P$}\put(14,2){$Q$}\put(30,2){$R$} \put(60,23){$S_{A..BP^{\prime
}Q^{\prime }..R^{\prime }}G_{PP^{\prime }} G_{QQ^{\prime}}..G_{RR^{\prime
}}$} \end{picture}.
\end{equation}
By substituting (28) and (29) into (26), we can obtain 
\begin{equation}
\Gamma ^{(2)}[\phi ,G]=-\frac{1}{12}S_{AQR}[\phi ]S_{BSP}[\phi
]G_{AB}G_{PQ}G_{RS}+\frac{1}{8}S_{ABPQ}[\phi ]G_{AB}G_{PQ}=%
\begin{picture}(120,20) \put(5,0){$-\frac{1}{12}$} \put(40,2){\circle{16}}
\put(32,2){\line(1,0){16}} \put(60,0){+$\frac{1}{8}$}
\put(94,2){\circle{16}} \put(110,2){\circle{16}} \end{picture}.
\end{equation}
In order to consider (26) in case of $l=3$, we first note that 
\begin{eqnarray}
\frac{\delta ^{2}\Gamma ^{(2)}[\phi ,G]}{\delta \phi _{A}\delta \phi _{B}}%
\text{ } &=&\frac{\delta ^{2}}{\delta \phi _{A}\delta \phi _{B}}[-\frac{1}{12%
}\begin{picture}(20,20) \put(10,2){\circle{16}} \put(2,2){\line(1,0){16}}
\end{picture} +\frac{1}{8}\begin{picture}(40,20) \put(10,2){\circle{16}}
\put(26,2){\circle{16}} \end{picture} ]=\frac{\delta }{\delta \phi _{B}}[-%
\frac{1}{6}\begin{picture}(40,20) \put(5,0){$A$}
\put(15,0){\framebox(4,4){}} \put(25,2){\circle{16}}
\put(17,2){\line(1,0){16}} \end{picture} +\frac{1}{8}\begin{picture}(40,20)
\put(16,0){\framebox(4,4){}} \put(10,2){\circle{16}} \put(26,2){\circle{16}}
\put(16,10){$A$} \end{picture} ] \\
&=&-\frac{1}{6}\begin{picture}(45,20) \put(8,0){$AB$}
\put(25,0){\framebox(4,4){}} \put(35,2){\circle{16}}
\put(27,2){\line(1,0){16}} \end{picture} -\frac{1}{6}\begin{picture}(50,20)
\put(5,0){$A$} \put(38,0){$B$} \put(15,0){\framebox(4,4){}}
\put(31,0){\framebox(4,4){}} \put(25,2){\circle{16}}
\put(17,2){\line(1,0){16}} \end{picture} +\frac{1}{8}\begin{picture}(40,20)
\put(16,0){\framebox(4,4){}}\put(10,2){\circle{16}} \put(26,2){\circle{16}}
\put(12,10){$AB$} \end{picture} .  \nonumber
\end{eqnarray}
and 
\begin{equation}
\frac{\delta \Gamma ^{(2)}[\phi ,G]}{\delta G_{AB}}=-\frac{1}{4}%
\begin{picture}(45,20) \put(5,0){$A$}
\put(15,0){\framebox(4,4){}}\put(25,2){\circle{16}}
\put(31,0){\framebox(4,4){}}\put(35,0){$B$}\end{picture} +\frac{1}{4}%
\begin{picture}(40,20) \put(8,-18){$AB$}
\put(13,-8){\framebox(4,4){}}\put(15,2){\circle{16}} \end{picture} \text{ }
\end{equation}
From (23), we also obtain 
\begin{equation}
\gamma _{A,BC}^{(1)}=\frac{\delta ^{2}\Gamma ^{(2)}[\phi ,G]}{\delta \phi
_{A}\delta G_{BC}}=-\frac{1}{4}\begin{picture}(45,20) \put(0,0){$AB$}
\put(15,0){\framebox(4,4){}}\put(25,2){\circle{16}}
\put(31,0){\framebox(4,4){}}\put(35,0){$C$}\end{picture} -\frac{1}{4}%
\begin{picture}(50,20) \put(5,0){$B$}
\put(15,0){\framebox(4,4){}}\put(25,2){\circle{16}}
\put(31,0){\framebox(4,4){}}\put(35,0){$AC$}\end{picture} +\frac{1}{4}%
\begin{picture}(40,20) \put(3,-18){$ABC$}
\put(13,-8){\framebox(4,4){}}\put(15,2){\circle{16}} \end{picture} \text{ }
\end{equation}
and 
\begin{equation}
\Delta _{AB,CD}^{(1)}=-2\frac{\delta ^{2}\Gamma ^{(2)}[\phi ,G]}{\delta
G_{AB}\delta G_{CD}}=\frac{1}{2}\begin{picture}(55,20) \put(0,0){$AC$}
\put(15,0){\framebox(4,4){}}\put(18,2){\line(1,0){16}}
\put(32,0){\framebox(4,4){}}\put(38,0){$BD$}\end{picture} +\frac{1}{2}%
\begin{picture}(55,20) \put(0,0){$AD$}
\put(15,0){\framebox(4,4){}}\put(18,2){\line(1,0){16}}
\put(32,0){\framebox(4,4){}}\put(38,0){$BC$}\end{picture} -\frac{1}{2}%
\begin{picture}(40,20) \put(6,5){$ABCD$} \put(14,-2){\framebox(6,4){}}
\end{picture}
\end{equation}
From (35), we can obtain the order $\hbar $ term of the inverse of the $%
\Delta $ as 
\begin{equation}
\Delta _{AB,CD}^{-1(1)}=-\text{ }\Delta _{AB,PQ}^{-1(0)}\Delta
_{PQ,RS}^{(1)}\Delta _{RS,CD}^{-1(0)}=-\frac{1}{2}\begin{picture}(45,20)
\put(10,0){\line(1,0){20}} \put(10,-10){\line(0,1){20}}
\put(30,-10){\line(0,1){20}} \put(2,10){$A$}
\put(2,-15){$D$}\put(32,10){$B$} \put(32,-15){$C$}\end{picture} -\frac{1}{2}%
\begin{picture}(45,20) \put(10,0){\line(1,0){20}}
\put(10,-12){\line(0,1){20}} \put(30,-10){\line(0,1){20}} \put(2,10){$A$}
\put(2,-15){$C$}\put(32,10){$B$} \put(32,-15){$D$}\end{picture} +\frac{1}{2}%
\begin{picture}(55,20) \put(10,0){\line(1,0){20}}
\put(20,-10){\line(0,1){20}} \put(2,0){$A$} \put(18,12){$B$}\put(33,0){$C$}
\put(18,-20){$D$}\end{picture}
\end{equation}
In case of the three-loop effective action $\Gamma ^{(3)}$, (26) becomes 
\begin{equation}
\frac{\delta \Gamma ^{(3)}[\phi ,G]}{\delta G_{AB}}G_{AB}=\frac{1}{2}G_{AB}%
\frac{\delta ^{2}\Gamma ^{(2)}}{\delta \phi _{A}\delta \phi _{B}}%
+G_{AB}\gamma _{A,DE}^{(0)}\Delta _{DE,PQ}^{-1(1)}\gamma _{B,PQ}^{(0)}\text{ 
}+2G_{AB}\gamma _{A,DE}^{(1)}\Delta _{DE,PQ}^{-1(0)}\gamma _{B,PQ}^{(0)}%
\text{ }
\end{equation}
By substituting (24),(32),(34) and (36) into (37), we obtain 
\begin{equation}
\Gamma ^{(3)}[\phi ,G]=-\frac{1}{48}\begin{picture}(40,20)
\put(15,3){\circle{20}} \put(5,3){\line(2,1){10}} \put(5,3){\line(2,-1){10}}
\put(25,3){\line(-2,1){10}} \put(25,3){\line(-2,-1){10}} \end{picture} +%
\frac{1}{8}\begin{picture}(40,20) \put(15,3){\circle{20}}
\put(15,-7){\line(1,2){8}}\put(15,-7){\line(-1,2){8}} \end{picture} -\frac{1%
}{24}\begin{picture}(40,20) \put(15,3){\circle{20}}
\put(15,13){\line(0,-1){10}} \put(5,3){\line(1,0){20}} \end{picture}
-\frac{1}{12}\begin{picture}(60,20) \put(15,3){\circle{20}} \put(35,3){\circle{20}}
\put(5,3){\line(1,0){20}}\end{picture} +\frac{1}{48}\begin{picture}(40,20)
\put(12,0){\circle{20}} \put(32,0){\circle{20}} \put(22,3){\line(1,2){8}}
\put(22,3){\line(-1,2){8}} \put(14,20){\line(1,0){16}} \end{picture}\end{equation}
Finally, in case of the four-loop effective action $\Gamma ^{(4)}$,  (26) becomes
\begin{eqnarray}
\frac{\delta \Gamma ^{(4)}[\phi ,G]}{\delta G_{AB}}G_{AB} &=&[\frac{1}{2}%
\frac{\delta ^{2}\Gamma ^{(3)}}{\delta \phi _{A}\delta \phi _{B}}+2\gamma
_{A,PQ}^{(2)}\Delta _{PQ,RS}^{-1(0)}\gamma _{B,RS}^{(0)}+\gamma
_{A,PQ}^{(1)}\Delta _{PQ,RS}^{-1(0)}\gamma _{B,RS}^{(1)}  \nonumber \\
&&+2\gamma _{A,PQ}^{(1)}\Delta _{PQ,RS}^{-1(1)}\gamma _{B,RS}^{(0)}+\gamma
_{A,PQ}^{(0)}\Delta _{PQ,RS}^{-1(2)}\gamma _{B,RS}^{(0)}]G_{AB}.
\end{eqnarray}
Here $\gamma ^{(0)},\gamma ^{(1)}$ and $\Delta ^{-1(1)}$ is given in
(24),(34) and (36) and $\Delta ^{-1(2)}$ is given by 
\begin{equation}
\Delta ^{-1(2)}=-\Delta ^{-1(0)}\Delta ^{(2)}\Delta ^{-1(0)}+\Delta
^{-1(0)}\Delta ^{(1)}\Delta ^{-1(0)}\Delta ^{(1)}\Delta ^{-1(0)}
\end{equation}
In order to obtain the RHS of (39), we need to evaluate $\frac{\delta
^{2}\Gamma ^{(3)}}{\delta \phi _{A}\delta \phi _{B}},\gamma ^{(2)}$ and $%
\Delta ^{(2)}$ from $\Gamma ^{(3)}[\phi ,G]$. We have
considered only cubic and quartic vertices for simplicity and the results
are given in Table 1. By substituting the results of Table 1 to RHS of (39)
we obtain
\begin{eqnarray}
\Gamma ^{(4)}[\phi ,G]&=&
\frac{1}{8}
\begin{picture}(46,30) \put(20,10){\circle{30}} \put(27,10){\line(1,0){10}}
\put(20,-5){\line(1,2){12}}\put(20,-5){\line(-1,2){12}} \end{picture} 
-\frac{1}{16}
\begin{picture}(46,30) \put(20,10){\circle{30}} \put(10,0){\line(1,0){20}}
\put(10,0){\line(0,1){23}}\put(30,0){\line(0,1){23}} \end{picture}
-\frac{1}{8} 
\begin{picture}(46,30) \put(20,10){\circle{30}} \put(10,0){\line(1,1){22}}
\put(10,0){\line(0,1){23}}\put(30,0){\line(0,1){23}} \end{picture}
-\frac{1}{12}
\begin{picture}(46,30) \put(20,10){\circle{30}} \put(10,10){\line(1,0){20}}
\put(10,-2){\line(0,1){25}}\put(30,-2){\line(0,1){25}} \end{picture} \nonumber \\
&+&\frac{1}{48}
\begin{picture}(46,30) \put(20,10){\circle{30}} \put(10,0){\line(1,0){20}}
\put(10,0){\line(2,5){10}}\put(30,0){\line(-2,5){10}} \end{picture} 
+\frac{1}{48} 
\begin{picture}(46,30) \put(20,10){\circle{30}}
 \put(5,10){\line(1,0){30}}\put(20,-5){\line(0,1){30}} \end{picture} 
-\frac{1}{8}
\begin{picture}(46,30) \put(20,10){\circle{30}} \put(20,18){\line(0,-1){15}}
\put(5,10){\line(2,1){15}} \put(5,10){\line(2,-1){15}} 
\put(35,10){\line(-2,1){15}} \put(35,10){\line(-2,-1){15}}  \end{picture} 
-\frac{1}{72}
\begin{picture}(46,30) \put(20,10){\circle{30}} \put(10,3){\line(1,0){20}}
\put(10,-2){\line(0,1){5}}\put(30,-2){\line(0,1){5}}
\put(10,3){\line(5,4){22}} \put(30,3){\line(-1,2){5}}
\put(21,15){\line(-1,2){5}} \end{picture} 
\end{eqnarray}
which agrees with the previous result given in \cite{2PI}.

\section{ Discussions and Conclusions}

In this paper, we have obtained the recursion relation to obtain the Feynman
diagrams of the CJT effective action by using the functional derivative
identities. We have shown the two-particle-irreducibility of the CJT effective
action and have obtained a Feynman diagrams of the CJT effective action
up to the four-loop order. The result agrees with the results obtained previously
by extracting 2PI Feynman diagrams from the 1PI effective action. 
The extension of the method we have used in this paper to
the case of recursive generation of the Feynman diagrams of the
effective action  obtained from the higher order Legendre transformation of the
generating functional is in progress.

\begin{acknowledgements}
This research was supported in part by the Institute of Natural Science.
\end{acknowledgements}

\bibliographystyle{plain}
\bibliography{2PI}

\begin{center} TABLES \end{center}
\begin{tabular}{|l|l|lll|lll|}
\hline
& $\frac{1}{2}\frac{\delta ^{2}\Gamma ^{(3)}}{\delta \phi _{A}\delta \phi
_{B}}$ &  & $\gamma _{A,BC}^{(2)}$ $=\frac{\delta ^{2}\Gamma ^{(3)}}{\delta
\phi _{A}\delta G_{BC}}$ &  &  & $\Delta _{AB,CD}^{(2)}$ & $=-2\frac{\delta
^{2}\Gamma ^{(3)}}{\delta G_{AB}\delta G_{CD}}$ \\ \hline
\begin{picture}(40,25) \put(0,10){$- \frac{1}{48}$} \put(30,10){\circle{20}}
\put(20,10){\line(2,1){10}} \put(20,10){\line(2,-1){10}}
\put(40,10){\line(-2,1){10}} \put(40,10){\line(-2,-1){10}} \end{picture} & 
&  &  &  & \begin{picture}(55,25) \put(0,10){$ \frac{1}{4}$}
\put(37,10){\circle{20}} \put(15,13){$AC$} \put(45,13){$BD$}
\put(25,8){\framebox(4,4){}}\put(45,8){\framebox(4,4){}} \end{picture} & %
\begin{picture}(50,25) \put(0,10){$+\frac{1}{4}$} \put(37,10){\circle{20}}
\put(15,13){$AD$} \put(45,13){$BC$}
\put(25,8){\framebox(4,4){}}\put(45,8){\framebox(4,4){}} \end{picture} &  \\ 
\hline
\begin{picture}(40,25) \put(30,10){\circle{20}} \put(10,10){$ \frac{1}{8}$}
\put(30,0){\line(1,2){8}}\put(30,0){\line(-1,2){8}} \end{picture} & %
\begin{picture}(40,25) \put(5,10){$ \frac{1}{8}$} \put(30,10){\circle{20}}
\put(30,0){\line(1,2){8}}\put(30,0){\line(-1,2){8}} \put(13,16){$A$}
\put(42,16){$B$} \put(20,13){\framebox(4,4){}}\put(35,13){\framebox(4,4){}}
\end{picture} & \begin{picture}(70,25) \put(0,10){$ \frac{1}{8}$}
\put(30,10){\circle{20}} \put(5,10){$AC$} \put(50,10){\circle{20}}
\put(65,10){$B$} \put(18,8){\framebox(4,4){}} \put(58,8){\framebox(4,4){}}
\end{picture} & \begin{picture}(80,25) \put(0,10){$+\frac{1}{8}$}
\put(38,10){\circle{20}} \put(13,10){$BC$} \put(58,10){\circle{20}}
\put(73,10){$A$} \put(25,8){\framebox(4,4){}} \put(65,8){\framebox(4,4){}}
\end{picture} & \begin{picture}(55,30) \put(0,10){$- \frac{1}{4}$}
\put(35,10){\circle{20}} \put(25,10){\line(1,0){20}} \put(15,10){$A$}
\put(30,24){$BC$} \put(23,8){\framebox(4,4){}} \put(33,18){\framebox(4,4){}}
\end{picture} & \begin{picture}(55,25) \put(0,10){$-\frac{1}{4}$}
\put(18,10){\line(1,0){10}} \put(26,8){\framebox(4,4){}} \put(13,15){$AC$}
\put(38,10){\circle{20}} \put(53,10){$B$} \put(31,10){$D$}
\put(16,8){\framebox(4,4){}} \put(46,8){\framebox(4,4){}} \end{picture} & %
\begin{picture}(50,25) \put(0,10){$-\frac{1}{4}$}
\put(21,10){\line(1,0){10}} \put(29,8){\framebox(4,4){}} \put(13,15){$AD$}
\put(41,10){\circle{20}} \put(53,13){$B$} \put(34,10){$C$}
\put(19,8){\framebox(4,4){}} \put(49,8){\framebox(4,4){}} \end{picture} & %
\begin{picture}(55,25) \put(0,10){$-\frac{1}{4}$}
\put(18,10){\line(1,0){10}} \put(26,8){\framebox(4,4){}} \put(12,15){$AC$}
\put(38,10){\circle{20}} \put(50,13){$D$} \put(31,10){$B$}
\put(16,8){\framebox(4,4){}} \put(46,8){\framebox(4,4){}} \end{picture} \\ 
&  & \begin{picture}(55,30) \put(0,10){$- \frac{1}{4}$}
\put(35,10){\circle{20}} \put(25,10){\line(1,0){20}} \put(15,10){$B$}
\put(30,24){$AC$} \put(23,8){\framebox(4,4){}} \put(33,18){\framebox(4,4){}}
\end{picture} & \begin{picture}(60,30) \put(0,10){$- \frac{1}{4}$}
\put(35,10){\circle{20}} \put(43,8){\framebox(4,4){}} \put(45,15){$C$}
\put(25,10){\line(1,0){20}} \put(15,10){$A$} \put(30,24){$B$}
\put(23,8){\framebox(4,4){}} \put(33,18){\framebox(4,4){}} \end{picture} & %
\begin{picture}(55,30) \put(0,10){$- \frac{1}{4}$} \put(35,10){\circle{20}}
\put(43,8){\framebox(4,4){}} \put(45,15){$C$} \put(25,10){\line(1,0){20}}
\put(15,10){$B$} \put(30,24){$A$} \put(23,8){\framebox(4,4){}}
\put(33,18){\framebox(4,4){}} \end{picture} & \begin{picture}(55,25)
\put(0,10){$-\frac{1}{4}$} \put(18,10){\line(1,0){10}}
\put(26,8){\framebox(4,4){}} \put(10,15){$AD$} \put(38,10){\circle{20}}
\put(50,13){$C$} \put(31,10){$B$} \put(16,8){\framebox(4,4){}}
\put(46,8){\framebox(4,4){}} \end{picture} & \begin{picture}(50,25)
\put(0,10){$-\frac{1}{4}$} \put(18,10){\line(1,0){10}} \put(15,15){$AC$}
\put(38,10){\circle{20}} \put(45,13){$BD$} \put(18,8){\framebox(4,4){}}
\put(46,8){\framebox(4,4){}} \end{picture} & \begin{picture}(55,25)
\put(0,10){$-\frac{1}{4}$} \put(18,10){\line(1,0){10}} \put(15,15){$AD$}
\put(38,10){\circle{20}} \put(50,13){$BC$} \put(16,8){\framebox(4,4){}}
\put(46,8){\framebox(4,4){}} \end{picture} \\ 
&  &  &  &  & \begin{picture}(55,30) \put(0,10){$-\frac{1}{2}$}
\put(36,18){\framebox(4,4){}} \put(38,10){\circle{20}} \put(48,10){$D$}
\put(36,25){$B$} \put(13,10){$AC$} \put(26,8){\framebox(4,4){}}
\put(46,8){\framebox(4,4){}} \end{picture} & \begin{picture}(55,30)
\put(0,10){$-\frac{1}{2}$} \put(36,18){\framebox(4,4){}}
\put(38,10){\circle{20}} \put(48,10){$C$} \put(36,25){$B$} \put(13,10){$AD$}
\put(26,8){\framebox(4,4){}} \put(46,8){\framebox(4,4){}} \end{picture} & %
\begin{picture}(55,20)\put(5,10){$+[A \longleftrightarrow B]$} \end{picture}
\\ \hline
\begin{picture}(40,25) \put(0,10){$- \frac{1}{24}$} \put(30,10){\circle{20}}
\put(30,20){\line(0,-1){10}} \put(20,10){\line(1,0){20}} \end{picture} & %
\begin{picture}(50,25) \put(0,10){$- \frac{1}{4}$} \put(35,10){\circle{20}}
\put(35,20){\line(0,-1){10}} \put(25,10){\line(1,0){20}} \put(15,10){$A$}
\put(45,10){$B$} \put(23,8){\framebox(4,4){}} \put(43,8){\framebox(4,4){}}
\end{picture} & \begin{picture}(50,25) \put(0,10){$- \frac{1}{4}$}
\put(35,10){\circle{20}} \put(35,20){\line(0,-1){20}} \put(15,10){$A$}
\put(45,10){$BC$} \put(23,8){\framebox(4,4){}} \put(43,8){\framebox(4,4){}}
\end{picture} & \begin{picture}(50,25) \put(0,10){$- \frac{1}{4}$}
\put(35,10){\circle{20}} \put(35,20){\line(0,-1){20}} \put(15,10){$B$}
\put(45,10){$AC$} \put(23,8){\framebox(4,4){}} \put(43,8) {\framebox(4,4){}}
\end{picture} & \begin{picture}(50,30) \put(0,10){$- \frac{1}{2}$}
\put(35,10){\circle{20}} \put(33,18){\framebox(4,4){}} \put(35,22){$C$}
\put(35,20){\line(0,-1){20}} \put(15,10){$A$} \put(45,10){$B$}
\put(23,8){\framebox(4,4){}} \put(43,8) {\framebox(4,4){}} \end{picture} & %
\begin{picture}(50,25) \put(0,10){$\frac{1}{2}$} \put(18,10){\line(1,0){10}}
\put(36,18){\framebox(4,4){}} \put(10,15){$AC$} \put(38,10){\circle{20}}
\put(50,13){$B$} \put(36,22){$D$} \put(16,8){\framebox(4,4){}}
\put(46,8){\framebox(4,4){}} \end{picture} & \begin{picture}(55,25)
\put(0,10){$+\frac{1}{2}$} \put(18,10){\line(1,0){10}}
\put(36,18){\framebox(4,4){}} \put(15,15){$AD$} \put(38,10){\circle{20}}
\put(50,13){$B$} \put(36,22){$C$} \put(16,8){\framebox(4,4){}}
\put(46,8){\framebox(4,4){}} \end{picture} & \begin{picture}(55,25)
\put(0,10){$+\frac{1}{2}$} \put(18,10){\line(1,0){10}}
\put(36,18){\framebox(4,4){}} \put(15,15){$BC$} \put(38,10){\circle{20}}
\put(50,13){$A$} \put(36,22){$D$} \put(16,8){\framebox(4,4){}}
\put(46,8){\framebox(4,4){}} \end{picture} \\ 
&  &  &  &  & \begin{picture}(55,25) \put(0,10){$+\frac{1}{2}$}
\put(18,10){\line(1,0){10}} \put(36,18){\framebox(4,4){}} \put(10,15){$BD$}
\put(38,10){\circle{20}} \put(50,13){$A$} \put(36,22){$C$}
\put(16,8){\framebox(4,4){}} \put(46,8){\framebox(4,4){}} \end{picture} & %
\begin{picture}(50,30) \put(0,12){$+\frac{1}{2}$}
\put(35,20){\framebox(4,4){}} \put(37,12){\circle{20}} \put(50,10){$D$}
\put(35,25){$A$} \put(17,10){$C$} \put(33,6){$B$}
\put(35,0){\framebox(4,4){}} \put(25,10){\framebox(4,4){}}
\put(45,10){\framebox(4,4){}} \end{picture} &  \\ \hline
\end{tabular}

TABLE 1. Contributions from each diagram of $\Gamma ^{(3)}[\phi ,G]$ ( first
column ) to $\frac{1}{2}\frac{\delta ^{2}\Gamma ^{(3)}}{\delta \phi
_{A}\delta \phi _{B}},\gamma ^{(2)}$ and $\Delta ^{(2)}.$
\newline
\newline
\newline

\newpage
\begin{center} FIGURE CAPTIONS \end{center}
\begin{picture}(500,80) 

\put(142,18){\oval(46,30)}  

\put(182,3){\framebox(50,30)}

\put(272,18){\oval(46,30)}  
\put(130,14){$\gamma_{A,PQ}$} \put(260,14){$\gamma_{B,RS}$}
\put(190,14){$\Delta_{PQ,RS}^{-1}$} \put(200,52){$G_{AB}$}

\put(164,10){\line(1,0){20}}  \put(232,10){\line(1,0){20}}
\put(162,30){\line(1,0){20}} \put(232,30){\line(1,0){22}}

\put(134,34){\line(0,1){12}} \put(282,34){\line(0,1){12}} 
\put(134,46){\line(1,0){148}} 

\end{picture}

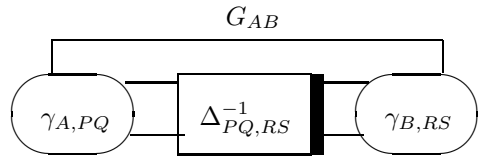
Fig.1. Graphical representation of second term of RHS of (26). Here, a line
represents the propagator $G_{AB}$.
\newline
\newline
\newline
\newline

\begin{picture}(500,50) 
\put(215,20){\oval(16,30)} \put(250,20){\oval(16,30)}
\put(222,20){\line(1,0){20}} 
\put(212,26){A} \put(212,6){B} \put(246,26){C} \put(246,6){D}
\end{picture}
\newline
\newline
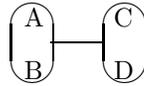
Fig.2. Type of a diagram which is not contained in $\Delta _{AB,CD}$ and $\Delta
_{AB,CD}^{-1}$.

\end{document}